\documentclass[prl,twocolumn,showpacs,preprintnumbers,superscriptaddress]{revtex4}
\usepackage{graphicx}
\usepackage{amsmath}
\usepackage{amssymb}
\usepackage{color}
\usepackage{dcolumn}
\usepackage{epsfig}
\usepackage{bm}

\newcommand{\lsim}{\mbox{\raisebox{-0.1cm}{$\;
\stackrel{\textstyle<}{\sim}\;$}}}

\begin{document}
%
\preprint{PREPRINT (\today)}


\title{Tuning topological disorder in MgB$_{2}$}

\author{D. Di Castro}
\affiliation{Coherentia-INFM-CNR and Physics Department, University of
Rome ``La Sapienza'', P.le A. Moro 5, I-00185, Rome, Italy}

\author{E. Cappelluti}
\affiliation{SMC-Institute for Complex System, INFM-CNR,
v. dei Taurini 19, 00185 Rome, Italy}
\affiliation{Physics Department, University of
Rome ``La Sapienza'', P.le A. Moro 5, I-00185, Rome, Italy}

\author{M. Lavagnini}
\affiliation{Physics Department, University of
Rome ``La Sapienza'', P.le A. Moro 5, I-00185, Rome, Italy}

\author{A. Sacchetti}
\affiliation{Coherentia-INFM-CNR and Physics Department, University of
Rome ``La Sapienza'', P.le A. Moro 5, I-00185, Rome, Italy}

\author{A.Palenzona}
\affiliation{INFM-LAMIA, Dipartimento di Fisica, Via Dodecaneso
33, 16146 Genova, Italy}

\author{M. Putti}
\affiliation{INFM-LAMIA, Dipartimento di Fisica, Via Dodecaneso
33, 16146 Genova, Italy}

\author{P. Postorino}
\affiliation{Coherentia-INFM-CNR and Physics Department, University of
Rome ``La Sapienza'', P.le A. Moro 5, I-00185, Rome, Italy}

\begin{abstract}
We carried out Raman measurements on neutron-irradiated and Al-doped MgB$_2$ samples. The irradiation-induced
topological disorder causes an unexpected appearance of high frequency spectral structures, similar to those
observed in lightly Al-doped samples. Our results show that disorder-induced violations of the selection rules
are responsible for the modification of the Raman spectrum in both irradiated and Al-doped samples. Theoretical
calculations of the phonon density of states support this hypothesis, and demonstrate that the high frequency
structures arise mostly from contributions at ${\bf q}\not=0$ of the E$_{2g}$ phonon mode.

\end{abstract}


\maketitle

A few years after the discovery of superconductivity in MgB$_2$, many of its characteristic properties have been
investigated and mostly clarified. Nowadays a wide consensus exists about MgB$_2$ being a phonon-mediated
two-gap/two-band superconductor \cite{Budko,Gonnelli,Souma,Liu,Choi}, where superconductivity mainly arises from
the coupling of the holes in the $\sigma$ band with the E$_{2g}$ phonon mode, which involves the in-plane
stretching vibrations of boron atoms \cite{Kortus,An,Kong,Yildirim}. In order to identify the fundamental
mechanisms of the superconducting pairing, a key role has been played by a variety of experimental techniques
which permitted the tuning of the electronic and lattice properties of MgB$_2$, such as the application of
external pressure \cite{Goncharov,DiCastroHP}, irradiation-induced disorder \cite{Bugoslavsky,Eisterer,Putti},
chemical substitution (e.g. Al on the Mg-site \cite{DiCastroTc,Renker} and C on the B-site
\cite{Carbon1,Carbon2}). In many of these cases, a reduction of the critical temperature, or even a suppression
of the superconducting phase, have been observed. In doped MgB$_2$ compounds, in particular, the debate focuses
on the relevance and the relative weight of two different microscopic mechanisms responsible for the reduction
of $T_c$, namely the band filling effects which tune the electron density of states \cite{Pena,Profeta,Kortus05}
and the disorder-induced increase of the inter-band scattering \cite{Erwin,Monni}.

Since the early investigations on MgB$_2$, Raman spectroscopy has been widely applied because the doubly
degenerate $E_{2g}$ mode, responsible for superconductivity, is the only Raman active among the four optical
phonons (A$_{2u}$, B$_{1g}$, E$_{1u}$ and E$_{2g}$)\cite{Yildirim,Kunc}. Indeed, the most prominent feature of
the measured Raman spectrum, namely the very broad peak (width $\sim$ 300 cm$^{-1}$) centered at around 620
cm$^{-1}$ \cite{Kunc,Postorino,Goncharov,Bohnen}), has been generally identified with the $E_{2g}$ phonon peak,
whose predicted frequency at the zone center $\Gamma$ lies within the 510-660 cm$^{-1}$ range
\cite{Kortus,Satta}. Despite the almost general consensus on the above assignment, however, there remains
unsolved controversies \cite{Kunc,Chen,Calandra}. Most of the doubts arises from the anomalously large phonon
linewidth and from its remarkable temperature dependence  \cite{Kunc,Postorino,Bohnen,Martinho}. Different
mechanisms, based on the strong electron-phonon (e-ph) coupling and on anharmonic effects, have been invoked to
explain these features \cite{Postorino,Goncharov,Bohnen,Cappelluti,FilmXiC}, although recent theoretical
calculations suggest these effects do not fully account for the experimental findings \cite{Calandra}. As an
alternative idea, the anomalously large lineshape was proposed to be related to the excitations of phonons
outside the $\Gamma$ point activated by lattice defects and multiphonon scattering \cite{Calandra}. This
hypothesis has also been suggested by some experimental works where the large phonon line-shape \cite{Chen}, and
other weak spectral features in the Raman spectrum \cite{Postorino}, were ascribed to phonon contributions,
throughout the Brillouin zone, activated by disorder-induced relaxation of the Raman selection rules.

Raman experiments have been also carried out on pure and doped MgB$_2$
films \cite{FilmYates,FilmXiSt,FilmXiC}
and on bulk chemically substituted MgB$_2$
compounds \cite{Carbon1,Carbon2,Postorino,Renker}.
Also in these cases,
the attempts at a quantitative
analysis lead to ambiguous and sometime contradictory results.
Contrary to the prediction of one Raman-active mode only,
the introduction of even small quantities of substitutional impurities appears to
drive the onset of additional high-frequency (HF) structures around
750 and 850 cm$^{-1}$ \cite{Postorino,Renker,Carbon2}, which are
absent or vanishingly small in the pure compound.
At higher levels of substitution, the Raman signal
further complicates, progressively approaching the spectrum of the end compound.
In Mg$_{1-x}$Al$_x$B$_2$ a two-steps behaviour is
observed: the HF structures are well detectable at $x=0.05$ and,
up to $x=0.25$, their intensity increases while keeping
almost constant the frequency positions; for $x>0.25$
the intensities keep growing but the
frequencies increase and the lineshapes narrow, evolving
as a whole towards the spectrum of the parent compound
AlB$_2$ \cite{Postorino,Renker}.
However, since C and Al doping induces, at the same time,
charge doping, volume compression and on-site lattice disorder,
it is particularly difficult to disentagle their effects
on the Raman spectrum and on the superconducting properties.

The present paper aims to single out the effect of topological disorder by carring out a Raman study of
differently neutron irradiated MgB$_2$ samples \cite{Putti}. We show that neutron irradiation causes the
appearance of HF spectral structures, similar to those observed in lightly Al-doped samples. The comparison of
the present spectra with the literature data allows us to ascribe the unexpected HF Raman features, present in
the irradiated and lightly Al-doped sample, to a disorder-induced violation of the Raman selection rules.
Theoretical calculations of the phonon dispersion curves support this hypothesis and allow us to compare the
Raman results with differently ${\bf q}$-integrated phonon density of states (PDOS).

Isotopically enriched ($^{11}$B) polycrystalline MgB$_2$ was prepared by direct synthesis from pure elements
Three identical samples were irradiated by means of thermal neutrons at the source SINQ (Paul
Sherrer Institut, Switzerland) \cite{Putti}. Different irradiation times were applied to the samples to
accomplish 7.6*10$^{17}$, 1.0*10$^{19}$, and 3.9*10$^{19}$ n/cm$^{2}$. The most important damage mechanism in
our samples is driven by the neutron capture reactions by the residual amount of $^{10}$B present (less than
0.5$\%$). Lattice defects are created by the recoil of $^{4}$He and $^{7}$Li, produced by the nuclear reaction,
which are emitted isotropically. This makes the defect distribution very homogeneous. After the irradiation, the
resistivity progressively increases (16, 64, 124 $\mu \Omega$ cm) and $T_c$ decreases (35.9 K, 24.3 K, and 12.2
K) on increasing the fluence. For sake of comparison low Al-doped samples (5$\%$, 10$\%$, 20$\%$, 30$\%$ with a
$T_c$ of 36.6 K, 33.4 K, 29.1 K and 24.1 K) were also prepared following a similar procedure \cite{PuttiAl}.

Raman spectra were measured in back-scattering geometry,
using a confocal micro-Raman spectrometer equipped with
a charge coupled device detector and notch filter to reject
the elastically scattered light. The sample was
excited by the 632.8 nm line of a 30 mW He-Ne Laser.
Raman spectra were collected using a 20x objective (laser
spot about 10 $\mu$m$^2$ wide at the sample surface)
over the frequency range 200-1100 cm$^{-1}$, which includes
the whole phonon spectral region.
For each sample Raman spectra were collected from different points on the
sample surface and then averaged to avoid possible effects
of preferred microcrystalline orientation.

\begin{figure}[t]
\includegraphics[width=0.75\linewidth]{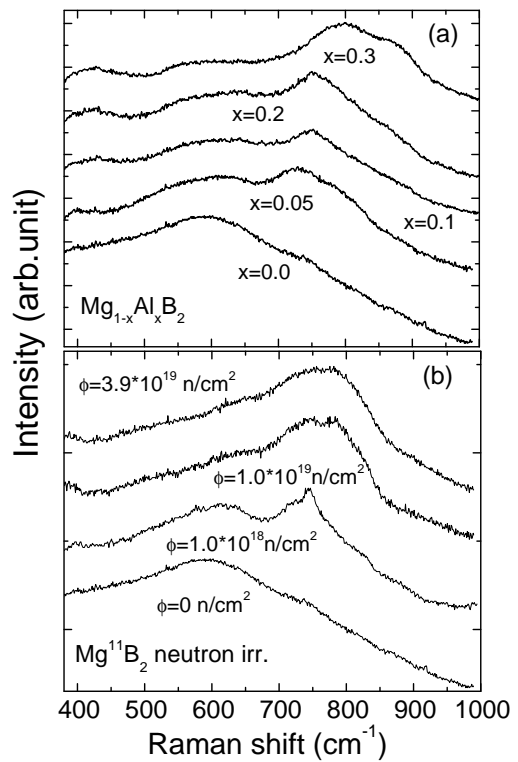}
\caption{Background subtracted Raman spectra of Al-doped (a) and neutron irradiated (b) MgB$_2$ samples. Spectra
are shifted upwards of a constant value.}
 \label{IrrAl}
\end{figure}

The Raman spectra of the irradiated and Al-doped samples are shown
in Fig.\ref{IrrAl}. Accordingly to literature data \cite{Postorino,Renker},
the effect of Al doping on the Raman signal is that of driving the onset of a HF spectral structure.
The Raman spectra of the irradiated samples, also shown in Fig.\ref{IrrAl},
present an unexpected strong dependence on the fluence, here observed for the first time.
Quite interestingly, the two series are surprisingly
similar: in both cases, HF structures appear
and become dominant at highest values of Al doping or
neutron fluence. The same features, occurring at about the same frequencies,
have been observed also in light C-doped MgB$_2$
\cite{Carbon2,FilmXiC} and disordered films \cite{FilmYates}.
Since neutron irradiation has the \textit{only} effect
of disordering the lattice, these results suggest that the HF structures
in the chemically substituted samples are produced mainly
by the lattice disorder as well.
In this context, the weak hardening of the frequencies
of the HF structures, well detectable only for 30$\%$ Al-doped samples,
could be likely ascribed to the Al-driven lattice compression.

In light of these results, and because of the unclear origin of the
HF structure, the core questions arise: what is the origin of the
HF structure and what is the role of the topological disorder
to make them visible? And, most importantly, is the ${\bf q}=0$ E$_{2g}$ phonon mode the one really probed in the
Al-doped and irradiated samples?
As we are going to discuss, we believe that the answer to the last question
is {\em negative} and that we are actually probing some kind
of phonon density of states.

\begin{figure}[t]
\includegraphics[width=0.8\linewidth]{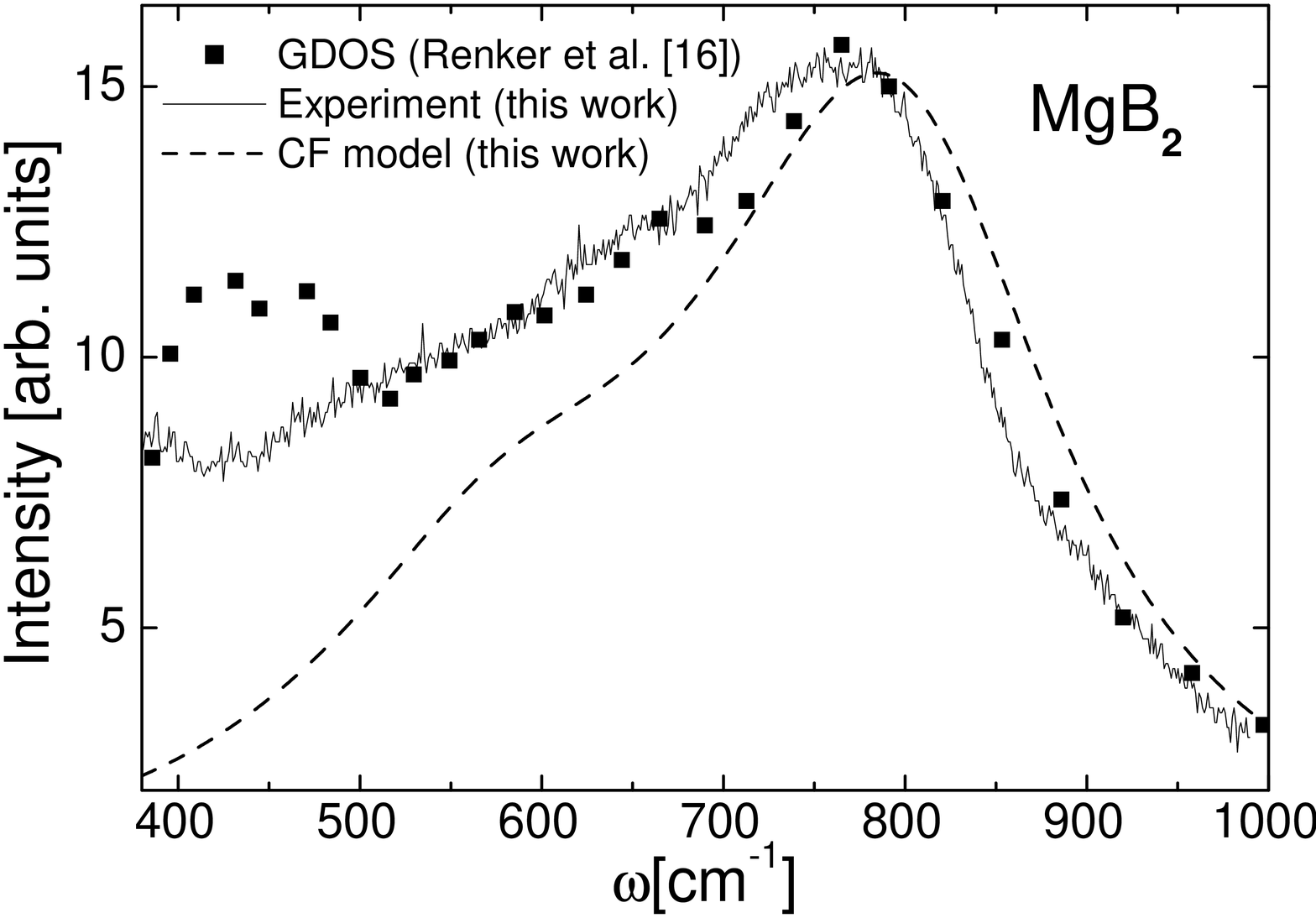}
\caption{Raman spectrum of the most irradiated sample
(full line) compared with the PDOS from the neutron experiment of Ref. \cite{Renker}(squares) and with the
calculated $E_{2g}$ projected PDOS (dashed line). All the reported quantities are empirically normalized to the
same peak intensity.} \label{DOS-Raman}
\end{figure}

In Fig. \ref{DOS-Raman} the Raman spectrum of the highest irradiated sample is compared with the PDOS of MgB$_2$
measured by inelastic neutron scattering \cite{Renker}.
 Since no selection rules apply to neutron scattering,
the excellent agreement between the experimental data over a wide frequency range (500-1000 cm$^{-1}$) suggests
that in disordered MgB$_2$, the Raman selection rules are notably relaxed and the Raman spectrum basically
reflects the PDOS. On the other side, the apparent disagreement below 500 cm$^{-1}$ suggests a remarkably
different weight of the phonon modes in the Raman and the neutron experiment. In the present case we can assume
that  the topological disorder, induced by neutron irradiation or by chemical substitution, mostly tunes the
violation of the Raman momentum selection rules but not of the eigenvector selection.

For a more quantitative check of this empirical hypothesis, we calculated the phonon dispersion of MgB$_2$
within a force-constant (FC)  shell model. Four elastic springs, specified by their tensor connecting B-Mg,
Mg-Mg first neighbors, and up to B-B second neighbors were considered. The corresponding twelve parameters were
obtained by fitting the first-principle calculated  phonon dispersion from Ref. \cite{Bohnen}. Since the FC
model is meant to reproduce the bare phonon dispersion, in the fitting procedure we did not include the $E_{2g}$
phonon branch along the $\Gamma- A$ direction, which is known to be strongly affected by the el-ph interaction.
The effects of the el-ph interaction involving the $E_{2g}$ boron mode with the electronic $\sigma$ band were
included through the self-energy renormalization of the phonon frequencies, by means of the two-dimensional
Lindhardt function, which, for the parabolic almost two-dimensional $\sigma$ bands, results to be a very good
approximation. Specific details, as well as the fitting values of the twelve FC parameters, can be found in Ref.
\cite{Campi}.

\begin{figure}[t]
\includegraphics[width=0.8\linewidth]{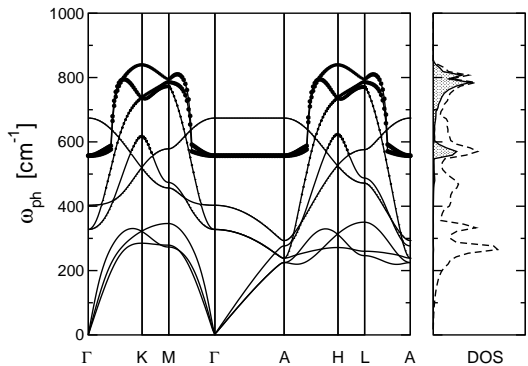}
\caption{Left panel: phonon dispersion of MgB$_2$ within the FC model. The thickness of the lines in the phonon
dispersion reflects the magnitude of the $E_{2g}$ component for each mode. Right panel: the corresponding total
PDOS (dashed line), and the  $E_{2g}$ projected PDOS (shaded area).} \label{Capp1}
\end{figure}

In Fig. \ref{Capp1} we show the so-calculated phonon dispersion curves $\omega_{{\bf q}, \mu}$ for the different
phonon branches $\mu$ and the total PDOS (dashed line). Our calculations enable to evaluate the $E_{2g}$
component of each phonon mode, $\eta_{\textbf{q}, \mu}^{E_{2g}} = |\hat\varepsilon_{\textbf{q}, \mu} \cdot
\hat\varepsilon_{E_{2g}}|^2$, which we believe to be effectively probed in our Raman measurements, by simply
projecting the generic phonon eigenvector $\hat\varepsilon_{\textbf{q}, \mu}$ on the $E_{2g}$ one,
$\hat\varepsilon_{E_{2g}}$. Thus, in Fig. \ref{Capp1} we show also the $E_{2g}$ character of each phonon branch,
as quantified by the thickness of the lines, as well as the $E_{2g}$-projected phonon densiy of states
($E_{2g}$-PDOS), shown as shaded area in the right panel. The resulting $E_{2g}$-PDOS, smeared by a Lorentzian
function with width $\Gamma= 100$ cm$^{-1}$, to simulate the finite phonon lifetime, is also shown in Fig.
\ref{DOS-Raman} for comparison purposes. The good agreement between the $E_{2g}$-PDOS and the Raman spectrum of
the highest irradiated sample supports the idea that the topological disorder induces a violation of the Raman
momentum selection rules, so that the effective spectra correspond to ${\bf q}$-integrated quantities. The
slight difference on the low frequency side could be ascribed to secondary contributions, disorder-activated,
from other Raman non-active modes, such as the $B_{1g}$. It is also worth to notice that the peak in the PDOS at
about 580-600 cm$^{-1}$ and the HF structures at $\sim$ 800 cm$^{-1}$, shown in Fig. \ref{Capp1}, arise from the
contribution of different regions of the Brillouin zone, respectively from the \textbf{q}-states close to the
$\Gamma$-$A$ and to K-M.

This observation suggests that
the onset of the HF structures in the disordered samples
could reflect the tuning of the breakdown of the momentum selection rules as
induced by the amount of disorder.
In this perspective we model the Raman intensity by
\begin{equation}
I(\omega) = \frac{1}{N_c}
\sum_{|\textbf{q}|\le q_c,\mu}
\eta_{\textbf{q}, \mu}^{E_{2g}}
\frac{1}{\pi} \frac{\Gamma_{q_c}}{(\omega -
\omega_{\textbf{q},\mu})^2 + \Gamma_{q_c}^2},
\label{eq2}
\end{equation}
where $N_c$ is the number of sampling points in the Brillouin zone
and where the momentum cut-off $q_c$ depends on the amount of disorder
and describes the \textbf{q}-space
effectively probed by the Raman measurements.
In particular, $q_c \rightarrow 0$ for ordered and pure compounds, thus selecting only
the Raman ${\bf q}=0$ signal, whereas for highly disordered samples
$q_c$ is larger than the size of the Brillouin zone and
the total ${\bf q}$-integrated PDOS is probed.
On a physical ground, we expect that also the phonon
scattering rate, $\Gamma_{q_c}$, is highly sensitive to the
amount of disorder, here parametrized by $q_c$, with
$\Gamma_{q_c}$ smallest for $q_c \rightarrow 0$ and
increasing with increasing $q_c$.
To take into account these effects we assumed a linear dependence on
$q_c$, namely $\Gamma_{q_c} = \alpha q_c$, where
$\alpha=120$ \AA/cm$^{-1}$.

\begin{figure}[t]
\includegraphics[width=0.85\linewidth]{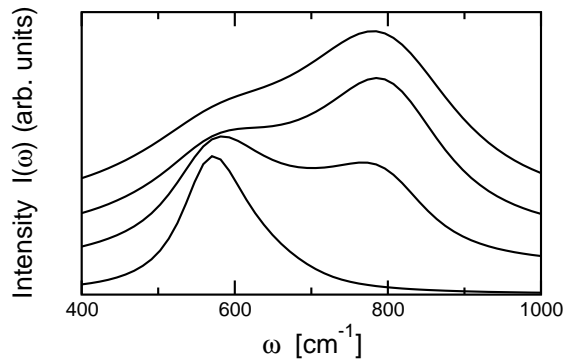}
\caption{Raman intensity from Eqs. (\ref{eq2}) for different values
of the cut-off $q_c$: (from the bottom to the top)
$q_c=0.30, 0,45, 0.60, 0.75$ {\AA}$^{-1}$. Curves are shifted upwards
for clarity.}
\label{Capp2}
\end{figure}

In Fig. \ref{Capp2} we show the intensities calculated using Eq. \ref{eq2} for different values of $q_c$ (0.30,
0,45, 0.60, 0.75 {\AA}$^{-1}$). For $|{\bf q}|\lsim 0.3$ {\AA}$^{-1}$ we are essentially probing only the flat
dispersion region close to the $\Gamma$ point shown in Fig. \ref{Capp1}, and only the low energy structure at
$\sim 600$ cm$^{-1}$ is visible. However, for larger $q_c$ the loss of the momentum selection rules is strong
enough to cause probing also of the $|{\bf q}|$ states close to the M and K points (we remind that $|{\bf
q}_{\rm M}|=0.39$ {\AA}$^{-1}$ and $|{\bf q}_{\rm K}|=0.68$ {\AA}$^{-1}$), making the HF structure of the
$E_{2g}$-PDOS of Fig. \ref{Capp1} rapidly visible. Finally, for $q_c$ larger than the size of the Brillouin
zone, all the $|{\bf q}|$ states are equally probed, and the signal $I(\omega)$ approaches the full
$E_{2g}$-PDOS (Fig. \ref{DOS-Raman}). The direct experimental {\em vs.} theoretical comparison in Fig.
\ref{DOS-Raman}, together with the remarkably similar trend shown by the spectra in Fig. \ref{IrrAl} and the
calculated intensities in Fig. \ref{Capp2}, reveal that, also in disordered systems, the Raman spectrum mostly
originates from the $E_{2g}$ phonon branch. Giving the key role of the $E_{2g}$ mode, this finding explains the
frequently observed correlation between  the Raman signal and the superconducting properties of MgB$_2$ based
compounds \cite{Renker,Carbon2,Postorino,FilmXiC,FilmYates}.

In conclusion, in this Letter we presented Raman spectra from differently neutron irradiated and Al-doped
MgB$_2$ samples. We have shown that, on increasing the neutron irradiation dose (i.e. topological disorder) as
well as Al-doping (i.e. topological disorder plus charge-doping), HF structures appear and rapidly become the
dominant spectral features. The remarkable similarity among our spectra and those reported in literature from
lightly C-doped samples suggests that the topological disorder, which is unavoidably present in chemically
substituted as well as in irradiated sample, is mostly responsible for the modifications of the Raman signal.
This is confirmed by the direct comparison between the Raman spectrum of the most irradiated sample and the
neutron PDOS \cite{Renker}. We also showed that the loss of the Raman momentum selection rules, tuned by the
amount of topological disorder, can explain in a natural way the onset of the HF structures. Thus the present
results provide a clear and unambiguous explanation of the Raman spectra in MgB$_2$, and  address Raman
spectroscopy as a very sensitive technique to investigate the role of topological disorder in MgB$_2$ and, more
in general, in  systems where a strong coupling between the lattice and the electronic degrees of freedom
exists. These findings also shed a new light on the interpretation of the previously reported Raman spectra in
MgB$_2$ based compounds.

\end{document}